\documentclass[12pt]{article}
\begin{document}
\title{Kinetic equations for thermal degradation of polymers}

\author{Aleksey D. Drozdov\footnote{E--mail:
Aleksey.Drozdov@mail.wvu.edu}\\
Department of Chemical Engineering\\
West Virginia University\\
P.O. Box 6102\\
Morgantown, WV 26506, USA}
\date{}
\maketitle

\begin{abstract}
Kinetic equations are analyzed for thermal degradation
of polymers.
The governing relations are based on the
fragmentation--annihilation concept.
Explicit solutions to these equations are derived in
two particular cases of interest.
For arbitrary values of adjustable parameters,
the evolution of the number-average and mass-average
molecular weights of polymers is analyzed numerically.
Good agreement is demonstrated between the results of numerical
simulation and experimental data.
It is revealed that the model can correctly predict observations
in thermo-gravimetric tests when its parameters are determined
by matching experimental data for the decrease in molecular
weight with exposure time.
\end{abstract}
\vspace*{10 mm}
\vspace*{10 mm}

\newpage

\section{Introduction}

This paper is concerned with the kinetics of thermal degradation
of polymers.
This subject has attracted substantial attention in the past half
a century both among the specialist in theoretical physics
and chemical engineering \cite{GS85,AE92}.
This may be explained by two reasons:
(i) scission (fragmentation) of macromolecules driven by
thermal fluctuations at elevated temperatures provides
a good example for the analysis of population dynamics
in complex systems,
and (ii) analysis of the degradation process becomes more and
more important due to an increase in the range of temperatures
for engineering applications,
recycling of post-consumer plastic waste,
as well as the use of polymers as biological implants and
matrices for drug delivery, where depolymerization is an inevitable
process affecting the life-time of an article.

Although the fact that binary scission of chains is
the main mechanism for thermal degradation of polymers
is widely accepted,
the applicability of the fragmentation concept to the analysis
of experimental data is rather limited.
This may be attributed to the fact that kinetic equations
for the ``linear" fragmentation process involve only one
adjustable parameter \cite{RBG98}, the number too small
to provide a reasonable approximation for observations.
The latter implies that generalizations of conventional
equations for the kinetics of fragmentation become a
meaningful subject for investigation.

The following ways for ``refinement" of the fragmentation
concept may be mentioned:
\begin{enumerate}
\item
An account for the influence of chains' length and
inhomogeneity of chains (the fact that the probability
of a breakage event depends on the position of a bond
along the backbone of a chain)
on the rate of scission \cite{ZM85,ZM86,ES93,SR95,TK00,KK01}.

\item
An introduction of two kinds of macromolecules in a network
(with strong and weak bonds) that have different
rates of fragmentation \cite{CB97,MCS97,KKM02}.

\item
An increase in the number of species in the population
balance laws by accounting for interactions between
chains and free radicals and the effect of the
concentration of free radicals on the scission
rate \cite{BBD94,BTV95,VV97,SKM01,KM02,MTW03}.
\end{enumerate}
Although these approaches seem rather attractive from
the theoretical viewpoint, they are grounded on
physical assumptions that are hard to be verified in
experiments, on the one hand, and they result in
overcomplicated explicit solutions (when the latter
can be developed), on the other.

Another way to make the fragmentation concept more flexible
for matching observations is to take into account some
physical processes at the micro-level that accompany
thermally-induced scission of macromolecules.
Two candidates for this role appear to be natural:
(i) aggregation (recombination) of broken chains
\cite{CMV95,JO00,LD03,YHW03},
and (ii) annihilation of fragmentation products
(either in the form of creation of inert species
that do not take part in further fragmentation
\cite{BK95,FR96,KB00}
or as diffusion of small-size fragmentation
products and their subsequent evaporation through
the surface of a specimen \cite{ECH90,HEL91,SR96,Enh03}).

In this study, we adopt the latter approach and accept
the conventional hypothesis that the diffusivity of
detached end- and side-groups at the degradation
temperature is so large that the kinetics of diffusion
may be disregarded.
This allows the number of adjustable parameters in
the governing equations to be reduced to four,
the number that gives an opportunity to analyze
the influence of all these quantities on the evolution of
molecular weight of polymers.
Analysis of the effect of diffusivity of oligomers
on the degradation kinetics will be the subject of
a subsequent paper.

Unlike previous studies, which focused mainly on the
scaling solutions to the fragmentation--annihilation
equations (the latter provide valuable information
about the properties of the degradation process at small
and large times), we concentrate on the kinetics of polymer
degradation within the time-scale of conventional tests.
Another difference between the present work and previous
publications is that we deal with changes in the
number-average and mass-average molecular weights,
the standard quantities measured in experiments.
This allows our analytical and numerical results
to be compared with available observations.

The objective of this study is two-fold:
\begin{itemize}
\item
To report some analytical and numerical solutions to
the fragmentation--annihilation equations.

\item
To establish restrictions on the material constants
in the governing relations imposed by the time--temperature
and time--molecular weight superposition principles,
and to find adjustable parameters in the kinetic equations
by matching observations.
\end{itemize}
We believe that the knowledge of these quantities
(or, at least, their orders of magnitude and mutual
relations between different constants) may be helpful for
further analytical and numerical investigation of the
degradation process.

In fitting experimental data, we concentrate on thermal
degradation of polystyrene (PS),
poly($\alpha$--methylstyrene) (PAMS)
and poly(L-lactide) (PLLA).
The choice of these polymers may be explained by their
wide use in industrial applications, including food service
packaging (PS and PAMS) and scaffolds for transplanted
organs (PLLA), the areas, where the degradation process is
of the highest importance.
Another (merely technical) reason for this choice is that
observations are available on these polymers with low
polydispersity indices (less than two), which implies
that no additional parameters associated with the initial
distribution of chains should be introduced into the model.

The paper is organized as follows.
We begin with the formulation of kinetic equations in
Section 2.
Some explicit solutions to these equations are derived
in Section 3.
The effect of material constants on the evolution of
molecular weight of polymers subjected to thermal
degradation is studied numerically in Section 4.
The time--molecular weight and time--temperature
superposition principles are discussed in Sections
5 and 6, respectively.
In Section 7, we demonstrate that the model can adequately
predict observations in thermo-gravimetric tests.
Some concluding remarks are formulated in Section 8.

\section{Kinetic equations}

In this section, kinetic equations are formulated for
the analysis of thermal degradation of polymers based
on the fragmentation--annihilation concept.

\subsection{Binary scission of chains}

Denote by $\bar{N}(t)$ the number of macromolecules per unit
mass of a polymer network at an arbitrary instant $t\geq 0$.
Following common practice, we treat chains as sequences
of segments connected by bonds.
Each chain in the network is entirely characterized
by the numbers of segments $k$ ($k=1,2,\ldots$).
Denote by $N_{k}(t)$ is the number of chains (per unit mass)
at time $t$ containing $k$ segments.
The functions $N_{k}(t)$ obey the conservation law
\begin{equation}
\bar{N}(t)=\sum_{k=1}^{\infty} N_{k}(t).
\end{equation}
Binary scission (fragmentation) of chains is described
by the reactions
\[
N_{k}\to N_{l}+N_{k-l} \qquad (l=1,\ldots, k-1).
\]
Denote by $\gamma$ the rate of scission (the number
of scission events per bond between segments per unit time).
Assuming $\gamma=\gamma_{0}$ to be independent of the
number of segments $k$, we arrive at the kinetic equations
for the functions $N_{k}(t)$
\begin{equation}
\frac{dN_{k}}{dt}(t) = -\gamma_{0} (k-1) N_{k}(t)
+2\gamma_{0} \sum_{j=k+1}^{\infty} N_{j}(t) .
\end{equation}
The coefficient $k-1$ in the first term describes
the number of possible scission events in a chain
containing $k$ segments.
The coefficient ``2" before the sum in Eq. (2) indicates
that there are two opportunities (``left" and ``right")
to obtain a chain with $k$ segments after scission of
a macromolecule with a larger number of segments.

Equation (2) can be easily generalized by assuming the rate
of scission $\gamma$ to be a function of chains' length.
Conventionally, the power-law dependence is adopted to
account for this dependence
\begin{equation}
\gamma(k)=\gamma_{0} k^{a},
\end{equation}
where $\gamma_{0}$ and $a$ are material parameters.
The kinetic equations for fragmentation of chains with
the length-dependent rate of scission (3) read
\begin{equation}
\frac{dN_{k}}{dt}(t) = -\gamma_{0}k^{a} (k-1) N_{k}(t)
+2\gamma _{0} \sum_{j=k+1}^{\infty} j^{a} N_{j}(t) .
\end{equation}

\subsection{Annihilation of chains}

Thermal fluctuations in a network induce not only
binary scission of macromolecules, but also detachment of
end- and side-groups from polymer chains.
As these groups are rather small, they have relatively
large diffusivity and can easily leave a specimen.
A decrease in a sample's mass with time driven by separation
of end- and side-groups and their desorption is treated
as their annihilation.

To make the model tractable from the mathematical standpoint,
we suppose that detachment of small groups within the interval
$[t, t+dt]$ may be thought of as transformation of a chain
with $k$ segments into a chain with $k-1$ segment.
Denote by $\Gamma$ the ratio of the number of chains that
lose a segment per unit time to the entire number of chains.
Assuming the annihilation rate to grow with the number of
segments in a chain following the power law similar to Eq. (3),
\begin{equation}
\Gamma(k)=\Gamma_{0}k^{b},
\end{equation}
where $\Gamma_{0}$ and $b$ are material parameters,
we arrive at the kinetic equations
\begin{equation}
\frac{dN_{k}}{dt}(t)=-\gamma_{0} k^{a} (k-1) N_{k}(t)
+2\gamma_{0} \sum_{j=k+1}^{\infty} j^{a} N_{j}(t)
+\Gamma_{0}\Bigl [ (k+1)^{b} N_{k+1}(t)-k^{b} N_{k}(t)\Bigr ] .
\end{equation}
The number-average molecular weight $M_{\rm n}$
and the mass-average molecular weight $M_{\rm w}$
of a polymer read
\begin{equation}
M_{\rm n}(t)=\frac{\sum_{k=1}^{\infty} kN_{k}(t)}
{\sum_{k=1}^{\infty} N_{k}(t)},
\qquad
M_{\rm w}(t)(t)=\frac{\sum_{k=1}^{\infty} k^{2} N_{k}(t)}
{\sum_{k=1}^{\infty} kN_{k}(t)}.
\end{equation}
The objective of this work is to analyze changes in
$M_{\rm n}(t)$ and $M_{\rm w}(t)$ with time $t$, when
the functions $N_{k}(t)$ are governed by Eq. (6).

\subsection{Transformation of the kinetic equations}

Introducing the concentrations of chains with $k$ segments,
\begin{equation}
n_{k}(t)=\frac{N_{k}(t)}{\bar{N}_{0}},
\end{equation}
where $\bar{N}_{0}=\bar{N}(0)$ is the total number of chains
at the initial instant $t=0$, we present Eq. (6) in the form
\begin{equation}
\frac{dn_{k}}{dt}(t) = -\gamma_{0} k^{a} (k-1) n_{k}(t)
+2\gamma_{0} \sum_{j=k+1}^{\infty} j^{a} n_{j}(t)
+\Gamma_{0}\Bigl [ (k+1)^{b} n_{k+1}(t)-k^{b} n_{k}(t)\Bigr ] .
\end{equation}
Following common practice, it is convenient to suppose
that the number of segments in a chain is large compared
to unity and to replace the discrete index $k$ in Eq. (9)
by a continuous argument $x$.
This results in the integro-differential
equation for the function $n(t,x)$
\begin{equation}
\frac{\partial n}{\partial t}(t,x) = -\gamma_{0} x^{a+1} n(t,x)
+2\gamma_{0} \int_{x}^{\infty} y^{a} n(t,y) dy
+\Gamma_{0} \frac{\partial}{\partial x}\Bigl (x^{b} n(t,x) \Bigr ).
\end{equation}
The initial condition for Eq. (10) reads
\begin{equation}
n(0,x)=n_{0}(x),
\end{equation}
where $n_{0}(x)$ is a given function.
We do not formulate boundary conditions for the function $n(t,x)$,
but assume that this function does not grow very strongly at $x=0$
and decays rapidly at $x\to \infty$ in the sense that the integrals
$M_{m}$ exist,
\begin{equation}
M_{m}(t)=\int_{0}^{\infty} x^{m} n(t,x) dx
\qquad
(m=0,1,2,\ldots).
\end{equation}

\section{Explicit solutions}

At $\Gamma_{0}=0$, i.e. when the annihilation process is suppressed,
an analytical solution of Eq. (10) was found in \cite{ZM86}
for an arbitrary integer $a\geq 0$.
However, an appropriate solution for $a>0$ is so complicated
that it can say practically nothing about the behavior of
the function $n(t,x)$ and its moments $M_{m}(t)$.
This implies that we confine ourselves to the case $a=0$
in the search for explicit solutions of Eq. (10).
Following \cite{ZM86}, we assume the parameter $b$ in Eq. (10)
to be an integer as well, $b=l$ ($l=1,2,\ldots$),
and re-write Eq. (10) in the form
\begin{equation}
\frac{\partial n}{\partial t}(t,x) = -\gamma_{0} n(t,x)
+2\gamma_{0} \int_{x}^{\infty} n(t,y) dy
+\Gamma_{0} \frac{\partial}{\partial x}\Bigl (x^{l} n(t,x) \Bigr ).
\end{equation}
Integrating Eq. (13) over $x^{m}$ ($m=0,1,\ldots$)
and using notation (12), we find that
\begin{equation}
\frac{dM_{m}}{dt}(t) = -\gamma_{0} M_{m+1}(t)
+2\gamma_{0} \int_{0}^{\infty} x^{m} dx \int_{x}^{\infty} n(t,y) dy
+\Gamma_{0} \int_{0}^{\infty} x^{m} \frac{\partial}{\partial x}
\Bigl (x^{l} n(t,x) \Bigr ) dx.
\end{equation}
The first integral in Eq. (14) is transformed by changing the order
of integration,
\[
\int_{0}^{\infty} x^{m} dx \int_{x}^{\infty} n(t,y) dy
=\int_{0}^{\infty} n(t,y)dy \int_{0}^{y} x^{m} dx
=\frac{1}{m+1} \int_{0}^{\infty} y^{m+1} n(t,y) dy
=\frac{1}{m+1} M_{m+1}(t) .
\]
The other integral is calculated by integration by parts,
\[
\int_{0}^{\infty} x^{m} \frac{\partial}{\partial x}
\Bigl (x^{l} n(t,x) \Bigr ) dx
=-m \int_{0}^{\infty} x^{m+l-1}n(t,x) dx=-mM_{m+l-1}(t).
\]
Substitution of these expressions into Eq. (14) results in
the differential equation
\begin{equation}
\frac{dM_{m}}{dt}(t)=-\gamma_{0} \biggl (1-\frac{2}{m+1}\biggr )
M_{m+1}(t)-\Gamma_{0} m M_{m+l-1}(t).
\end{equation}
Our purpose now is to find explicit solutions of Eq. (15) for
$l=1$ and $l=2$.
According to Eq. (5), the former case corresponds to scission
and annihilation of end-groups exclusively, whereas the latter
case describes annihilation of side-groups homogeneously
distributed along a chain's backbone.

\subsection{The case $l=1$}

Setting $l=1$ in Eq. (15), we obtain
\begin{equation}
\frac{dM_{m}}{dt}(t)=-\Gamma_{0} m M_{m}(t)
-\gamma_{0} \biggl (1-\frac{2}{m+1}\biggr )M_{m+1}(t).
\end{equation}
Introducing the notation
\begin{equation}
T_{m}(\tau)=M_{m}(t)\exp (\Gamma_{0} mt),
\qquad
\tau=\int_{0}^{t} \exp(-\Gamma_{0} s)ds,
\end{equation}
we transform Eq. (16) as follows:
\begin{equation}
\frac{d T_{m}}{d\tau}(\tau)=
-\gamma_{0} \biggl (1-\frac{2}{m+1}\biggr )T_{m+1}(\tau).
\end{equation}
Comparison of Eqs. (15) and (18) implies that in the special
case $l=1$,
the kinetics of the fragmentation--annihilation process
coincides with the kinetics of a mere fragmentation process
in the new time $\tau$ for the new moments $T_{m}$.
In particular, putting $m=0$ and $m=1$ in Eq. (18), we
find that
\[
\frac{dT_{0}}{d\tau}=\gamma_{0} T_{1},
\qquad
\frac{dT_{1}}{d\tau}=0.
\]
The solutions of these equations are given by
\begin{equation}
T_{0}(\tau)=T_{0}(0)+\gamma _{0} T_{1}(0)\tau,
\qquad
T_{1}(\tau)=T_{1}(0).
\end{equation}
The moment $T_{2}(\tau)$ is determined by the formula
\begin{equation}
T_{2}(\tau)=\frac{2}{(\gamma_{0} \tau)^{2}}
\Bigl [ \hat{n}_{0}(\gamma_{0} \tau)
-T_{0}(0)+T_{1}(0) \gamma_{0} \tau\Bigr ],
\end{equation}
where $\hat{n}_{0}(z)$ is the Laplace transform of the function
$n_{0}(x)$.
A detailed derivation of Eq. (20) is given in Appendix.
It follows from Eqs. (17), (19) and (20) that
\begin{eqnarray}
M_{0}(t) &=& \biggl [ M_{0}(0)+M_{1}(0) \frac{\gamma_{0}}{\Gamma_{0}}
\Bigl (1-\exp (-\Gamma_{0} t)\Bigr )\biggr ],
\nonumber\\
M_{1}(t) &=& M_{1}(0)\exp (-\Gamma_{0} t\Bigr ),
\nonumber\\
M_{2}(t) &=& 2 \Bigl (\frac{\Gamma_{0}}{\gamma_{0}}\Bigr )^{2}
\frac{\exp(-2\Gamma_{0} t)}{(1-\exp(-\Gamma_{0} t))^{2}}
\biggl [ \hat{n}_{0}\Bigl (\frac{\gamma_{0}}{\Gamma_{0}}
(1-\exp(-\Gamma_{0} t))\Bigr )
\nonumber\\
&&
-M_{0}(0)+M_{1}(0)\frac{\gamma_{0}}{\Gamma_{0}}
\Bigl (1-\exp(-\Gamma_{0} t)\Bigr ) \biggr ].
\end{eqnarray}
Substituting Eq. (21) into Eq. (7), where the sums are
replaced by integrals, we arrive at explicit expressions
for the number-average and mass-average molecular weights
as functions of time $t$.

\subsection{The case $l=2$}

Setting $l=2$ in Eq. (15), we obtain
\begin{equation}
\frac{dM_{m}}{dt}(t)=-\gamma_{0} \biggl (\frac{m-1}{m+1}
+\frac{\Gamma_{0}}{\gamma_{0}}m \biggr )M_{m+1}(t).
\end{equation}
Assuming the initial network of chains to be monodisperse,
\begin{equation}
n_{0}(x)=\delta (x-L),
\end{equation}
where $L$ is the initial length of chains,
we solve Eq. (22) by Charlesby's method \cite{Cha54}.
As Eq. (22) is linear with respect to the unknown function
$n(t,x)$, appropriate formulas for the moments $M_{m}(t)$
corresponding to an arbitrary initial condition $n_{0}(x)$
are developed by the superposition method.

The $m$th moment $M_{m}(t)$ is expanded into the Taylor
series in time,
\begin{equation}
M_{m}(t)=\sum_{k=0}^{\infty} \frac{M^{(k)}(0)}{k!}t^{k},
\end{equation}
where $M_{m}^{(k)}(0)$ stands for the $k$th derivative
at the point $t=0$.
Substitution of Eq. (24) into Eq. (22) implies that
for an arbitrary $k>1$,
\begin{equation}
M_{m}^{(k)}(0) = (-\gamma_{0})^{k} \prod_{j=0}^{k-1}
\biggl (\frac{m+j-1}{m+j+1}
+\frac{\Gamma_{0}}{\gamma_{0}}(m+j) \biggr )M_{m+k}(0).
\end{equation}
It follows from Eqs. (12) and (23) that
\begin{equation}
M_{m}(0)=L^{m}.
\end{equation}
Substitution of expressions (25) and (26) into Eq. (24)
results in
\begin{equation}
M_{m}(t)=L^{m}\biggl [ 1+\sum_{k=1}^{\infty} A_{mk} (-\gamma_{0} t L)^{k}
\biggr ],
\end{equation}
where
\[
A_{mk}=\frac{1}{k!}\prod_{j=0}^{k-1}
\biggl (\frac{m+j-1}{m+j+1}
+\frac{\Gamma_{0}}{\gamma_{0}}(m+j) \biggr ).
\]
Introducing the new variable $j^{\prime}=j+1$ and omitting the prime,
we obtain
\begin{equation}
A_{mk}
=\prod_{j=1}^{k} \frac{1}{j} \biggl (\frac{m+j-2}{m+j}
+\frac{\Gamma_{0}}{\gamma_{0}}(m+j-1) \biggr ).
\end{equation}
It follows from Eq. (27) that for an arbitrary initial
condition $n_{0}(x)$, the moments $M_{m}(t)$ are given by
\begin{eqnarray}
M_{m}(t) &=& \int_{0}^{\infty} n_{0}(x) x^{m}
\biggl [ 1+\sum_{k=1}^{\infty} A_{mk} (-\gamma_{0} t x)^{k}
\biggr ] dx
\nonumber\\
&=& M_{m}(0) + \sum_{k=1}^{\infty} A_{mk} M_{m+k}(0) (-\gamma_{0} t )^{k}.
\end{eqnarray}
Although Eq. (29) provides explicit expressions for the moments
$M_{m}(t)$, it is not rather convenient for the numerical analysis,
because the series converges slowly.
Formulas (28) and (29) are helpful, however,
for the evaluation of changes in $M_{m}(t)$ at small times,
when $\gamma_{0}t\ll 1$.
Neglecting terms beyond the first order of smallness in Eq. (29),
we find that
\begin{eqnarray}
&& M_{0}(t) = M_{0}(0)+M_{1}(0) \gamma_{0} t,
\qquad
M_{1}(t) = M_{1}(0)-M_{2}(0) \Gamma_{0} t,
\nonumber\\
&& M_{2}(t) = M_{2}(0)-M_{3}(0) \Bigl (\frac{\gamma_{0}}{3}
+2\Gamma_{0} \Bigr )t.
\end{eqnarray}
According to Eq. (30), changes in the moments $M_{0}(t)$
and $M_{1}(t)$ are governed by two different processes:
an increase in $M_{0}(t)$ is driven by fragmentation of
chains, whereas a decrease in $M_{1}(t)$ is induced by
annihilation of end- and side-groups.
Introducing the notation
\begin{equation}
d(t)=\frac{M_{1}(t)}{M_{1}(0)},
\qquad
d_{\rm n}(t)=\frac{M_{\rm n}(t)}{M_{\rm n}(0)},
\qquad
d_{\rm w}(t)=\frac{M_{\rm w}(t)}{M_{\rm w}(0)}
\end{equation}
and using Eq. (7), we present Eq. (30) in the form
\begin{eqnarray}
&& d(t)=1-M_{\rm w}(0)\Gamma_{0} t,
\qquad
d_{\rm n}(t)=1-\Bigl (M_{\rm n}(0)\gamma_{0}
+M_{\rm w}(0)\Gamma_{0} \Bigr ) t,
\nonumber\\
&& d_{\rm w}(t)=1-\biggl [ M_{\rm w}(0)\Gamma_{0}+M_{\rm z}(0)
\Bigl (\frac{\gamma_{0}}{3}+2\Gamma_{0} \Bigr )\biggr ] t,
\end{eqnarray}
where the conventional notation is used
$M_{\rm z}(t)=M_{3}(t)/M_{2}(t)$.

\section{Numerical analysis}

As explicit formulas for the moments $M_{m}(t)$ can be derived
only for a limited set of values of the exponents $a$ and $b$,
it is of interest to analyze the evolution of $M_{m}$ with time
$t$ numerically for arbitrary values of these parameters.
For this purpose, we integrate Eq. (9) numerically with the
initial condition
\begin{equation}
n_{k}(0)=\delta_{kK},
\end{equation}
where $\delta_{ij}$ denotes the Kronecker delta.
Equation (33) corresponds to the monodisperse distribution
of chains that contain $K$ segments at the initial
instant $t=0$.
To reduce the number of material parameters, we introduce
the dimensionless time $t^{\prime}=\gamma_{0}t$
and set
\begin{equation}
\eta=\frac{\Gamma_{0}}{\gamma_{0}}.
\end{equation}
In the new notation, Eq. (9) reads (the prime is omitted for
simplicity)
\begin{equation}
\frac{dn_{k}}{dt}(t) = - k^{a} (k-1) n_{k}(t)
+2 \sum_{j=k+1}^{K} j^{a} n_{j}(t)
+\eta \Bigl [ (k+1)^{b} n_{k+1}(t)-k^{b} n_{k}(t)\Bigr ]
\qquad
(k=1,2,\ldots,K).
\end{equation}
The limitation on the number of equations in Eq. (35) follows from
the fact that if the maximal number of segments in a chain
equals $K$ at $t=0$, no chains with higher number of segments
can appear at $t>0$ due to the fragmentation--annihilation
process.
Equations (33) and (35) involve four material constants:
$a$, $b$, $\eta$ and $K$.
In the numerical simulation, we fix $K=100$ and study the effect
of other parameters on the number-average and mass-average molecular
weights determined by Eqs. (7) and (8).
Integration of Eq. (35) is performed by the Runge--Kutta
method with the step $\Delta t=1.0\cdot 10^{-6}$.

First, we set $a=0$ and $b=1$ and study the effect of $\eta$
on the ratios $d_{\rm n}(t)$ and $d_{\rm w}(t)$ defined
by Eq. (31).
Figures 1 and 2 demonstrate that annihilation of end-
and side-groups affects the degradation process noticeably
when the rate of annihilation $\Gamma_{0}$ exceeds the rate
of scission $\gamma_{0}$ by an order of magnitude.

Afterwards, we fix $\eta=10.0$, $b=1.0$ and analyze
the influence of the exponent $a$ on the dimensionless
ratios $d_{\rm n}$ and $d_{\rm w}$.
Figures 3 and 4 reveal that an increase in $a$ results in
a very strong decrease in $d_{\rm n}$ and $d_{\rm w}$ for
any $t>0$.
When $a\approx 1$, the curves $d_{\rm n}(t)$ and $d_{\rm w}(t)$
steeply drop at small times and remain practically constant at
larger times.
These results make questionable the applicability of the
analytical solution developed in \cite{ZM86} (this solution
corresponds to $a=2$) for the description of the degradation
process.

Finally, we fix $a=0.2$, $\eta=10.0$ and assess the effect of
the exponent $b$ on $d_{\rm n}(t)$ and $d_{\rm w}(t)$.
Figures 5 and 6 show that the influence of $b$ is negligible
when $b\in [0,1)$.
However, for $b>1$, the effect of this parameter becomes
substantial, and an increase in $b$ causes a pronounced
decrease in number-average and mass-average molecular weights.

To ensure the accuracy of numerical simulation, we use three
tests.
First, we verify that at $\eta=0$, the first moment $M_{1}$
remains independent of time [this conclusion follows from
Eq. (15), where we set $\Gamma_{0}=0$].
Secondly, we increase $K$ by twice, decrease the rate of
fragmentation $\gamma_{0}$ by twice and check that the
moments $M_{\rm n}(t)$ and $M_{\rm w}(t)$ remain unchanged.
The latter implies that the results of numerical
analysis are independent of our choice of $K=100$.
Finally, we perform simulation with $a=0$ and $b=1$
and confirm that the numerical results for the moments
$M_{m}(t)$ ($m=0,1,2$) coincide with analytical solution (21).

\section{Time--molecular weight superposition principle}

Experimental data for monodisperse polymers reveal that
observations for the ratios of number-average and
mass-average molecular weights, $d_{\rm n}(t)$ and
$d_{\rm w}(t)$, obtained at various initial molecular
weights and plotted in semi-logarithmic coordinates
(versus the logarithm of time) may be superposed
(with an acceptable level of accuracy) by shifts along
the time axis.
The construction of a master-curve by shift of creep and
relaxation curves measured at various temperatures
is a conventional procedure in linear viscoelasticity
of polymers \cite{Fer80}.
The validity of this operation is based on the
time--temperature superposition principle, which asserts
that only the characteristic creep and relaxation times
are affected by temperature, while other parameters
(elastic moduli and shapes of the relaxation spectra)
remain temperature-independent.
Our aim now is to establish an analogous principle for
the effect of initial molecular weight of monodisperse
polymers on the fragmentation--annihilation process.
For this purpose, we return to integro-differential
equation (10) and introduce the dimensionless variables
\begin{equation}
t_{\ast}=\frac{t}{\Theta},
\qquad
x_{\ast}=\frac{x}{L},
\end{equation}
where $\Theta$ is the characteristic time for degradation.
In the new notation, Eq. (10) reads
\begin{equation}
\frac{\partial n}{\partial t_{\ast}}(t_{\ast},x_{\ast})
= \gamma_{0}\Theta L^{a+1}\biggl [- x_{\ast}^{a+1} n(t_{\ast},x_{\ast})
+2 \int_{x_{\ast}}^{\infty} y_{\ast}^{a} n(t_{\ast},y_{\ast}) dy_{\ast}
+\eta L^{b-a-2} \frac{\partial}{\partial x_{\ast}}
\Bigl (x_{\ast}^{b} n(t_{\ast},x_{\ast}) \Bigr )\biggr ].
\end{equation}
It follows Eq. (37) that any solution $n(t_{\ast},x_{\ast})$
is independent of the initial molecular weight $L$
[this statement implies the time--molecular weight superposition
principle] provided that
\begin{equation}
b=a+2.
\end{equation}
Under condition (38), the shift factor $A=\Theta/\Theta^{\rm ref}$,
where $\Theta^{\rm ref}$ is the characteristic time corresponding to
a reference molecular weight $L^{\rm ref}$, is determined
by the formula
\begin{equation}
\log A=(a+1)\log \frac{L}{L^{\rm ref}}
\end{equation}
with $\log=\log_{10}$.
According to the time--molecular weight superposition principle,
the number of adjustable parameters in the model may be
substantially reduced:
instead of four material constants, $\gamma_{0}$, $\Gamma_{0}$,
$a$ and $b$, we have only two parameters to be found, $\gamma_{0}$
and $\eta$: the exponent $a$ is uniquely determined from Eq. (39),
whereas the exponent $b$ is given by Eq. (38).

\subsection{Fitting of observations on polystyrene}

To find the values of $\gamma_{0}$ and $\eta$ and to assess
the influence of annihilation of end- and side-groups on the
degradation process,
we focus on observations reported by Madras et al.
\cite{MCS97} on thermal degradation of monodisperse PS
in a mineral oil at the temperature $T=275$~$^{\circ}$C.
For a detailed description of specimens and the experimental
procedure, the reader is referred to \cite{MCS97}.
The ratio $d_{\rm w}$ of mass-average molecular weights
is depicted versus exposure time $t$ in Figure 7.
The experimental data for $M_{\rm w}(0)=26$ kg/mol are presented
without changes.
Observations for the other initial molecular weights
($M_{\rm w}(0)=12$, 110, 210, 330 and 930 kg/mol)
are shifted along the time-axis by appropriate amounts
$A$ that are determined from the condition that all
available data produce a smooth master-curve.

The parameter $A$ is plotted versus the ratio of initial
molecular weights $M_{\rm w}(0)/M_{\rm w}^{\rm ref}(0)$
with $M_{\rm w}^{\rm ref}(0)=26$ kg/mol in Figure 8.
The experimental data are approximated by the function
\begin{equation}
\log A=A_{0}+A_{1}\log \frac{M_{\rm w}(0)}{M_{\rm w}^{\rm ref}(0)},
\end{equation}
where the coefficients $A_{k}$ ($k=0,1$) are determined
by the least-squares method.
Figure 8 demonstrates that Eq. (40) provides good matching
of the observations with $A_{0}\approx 0$ and $A_{1}\approx 1$.
With reference to this result,
we conclude from Eqs. (38), (39) and (40) that one should
set $a=0$ and $b=2$ in Eq. (35) in order to reproduce
the experimental data plotted in Figure 7.

The fact that $A_{1}\approx 1$, which implies that the
fragmentation rate $\gamma_{0}$ is proportional to the
initial mass-average molecular weight $M_{\rm w}(0)$,
\begin{equation}
\frac{\gamma_{0}}{\gamma_{0}^{\rm ref}}
=\frac{M_{\rm w}(0)}{M_{\rm w}^{\rm ref}(0)},
\end{equation}
has been revealed about 40 years ago \cite{BB65}.
It was also shown that Eq. (41) is valid when
the initial molecular weight $M_{\rm w}(0)$ is not very
large: for polymers with ultra-high molecular weights,
the fragmentation rate $\gamma_{0}$ grows like the square-root
of the molecular weight \cite{BW58}.

To find $\gamma_{0}$ and $\eta$, we fix some intervals
$[0,\gamma_{\max}]$ and $[0,\eta_{\max}]$,
where the ``best-fit" parameters $\gamma_{0}$ and
$\eta$ are assumed to be located,
and divide these intervals into $J$ subintervals by
the points $\gamma^{(i)}=i \Delta \gamma$,
and $\eta^{(j)}=j \Delta \eta$ ($i,j=1,\ldots,J-1$)
with $\Delta \gamma=\gamma_{\max}/J$ and
$\Delta \eta=\eta_{\max}/J$.
For any pair $\{ \gamma^{(i)}, \eta^{(j)} \}$,
Eq. (35) with initial condition (33) is integrated numerically
by the Runge--Kutta method with $K=100$ and
the time-step $\Delta t=0.1$.
This, relatively large, step is chosen because the fragmentation
rates under consideration $\gamma_{0}$ are quite small (of
order of 10$^{-5}$).
The best-fit parameters $\gamma_{0}$ and $\eta$ are determined
from the condition of minimum of the function
\[
R=\sum_{t_{m}} \Bigl [ d_{\rm w}^{\;\rm exp}(t_{m})
-d_{\rm w}^{\;\rm num}(t_{m}) \Bigr ]^{2},
\]
where the sum is calculated over all times $t_{m}$
at which observations are presented in Figure 7,
$d_{\rm w}^{\;\rm exp}$ is the ratio of mass-average
molecular weights measured in the tests,
and $d_{\rm w}^{\;\rm num}$ is given by Eqs. (7), (8) and (31).
Figure 7 demonstrates fair agreement between the
observations at various initial molecular weights and
the results of numerical simulation with
$\gamma_{0}=1.4\cdot 10^{-5}$ and $\eta=1.0$.
Two conclusions may be drawn from the numerical analysis:
(i) the model correctly describes the experimental data,
and (ii) the best-fit value $\eta=1.0$ found by matching
the observations on monodisperse polystyrene
results in a rather weak influence of the annihilation
process on the kinetics of thermal degradation (see Figures
1 and 2).
Our purpose now is to demonstrate that this is not the case
for other polymers.
To show that annihilation of end- and side-groups
noticeably affects the degradation process,
we analyze two sets of observations on PAMS
and PLLA obtained at various temperatures $T$.

\section{Time--temperature superposition principle}

We begin with the discussion of the effect of temperature $T$
on the kinetics of thermal degradation.
To establish conditions on the material constants which ensure
that the time--temperature superposition principle holds
for the ratios $d_{\rm n}$ and $d_{\rm w}$,
i.e. that the functions $d_{\rm n}(t)$ and $d_{\rm w}(t)$
measured at various temperatures and plotted in semi-logarithmic
coordinates may be shifted along the time-axis to construct a
master-curve \cite{CB97},
we return to the dimensionless integro-differential equation (37).

According to the Eyring theory \cite{Eyr36}, the effect of
temperature on the rates of thermally activated processes
may be accounted for by the formulas
\begin{equation}
\gamma_{0}=\gamma_{0}^{\rm ref}
\exp \biggl [ -\frac{E_{\gamma}}{R}
\Bigl ( \frac{1}{T}-\frac{1}{T^{\rm ref}} \Bigr )\biggr ],
\qquad
\Gamma_{0}=\Gamma_{0}^{\rm ref}
\exp \biggl [ -\frac{E_{\Gamma}}{R}
\Bigl ( \frac{1}{T}-\frac{1}{T^{\rm ref}} \Bigr )\biggr ],
\end{equation}
where $T$ is the absolute temperature,
$R$ is the universal gas constant,
$E_{\gamma}$ and $E_{\Gamma}$ are appropriate
activation energies,
and $\gamma_{0}^{\rm ref}$ and $\Gamma_{0}^{\rm ref}$ are
rates of fragmentation and annihilation at the reference
temperature $T^{\rm ref}$.
Equation (37) implies that the time--temperature superposition
principle is valid, provided that the exponents $a$ and $b$
are independent of temperature, while the rates of fragmentation
and annihilation, $\gamma_{0}$ and $\Gamma_{0}$, are affected by
$T$ in a similar fashion,
\begin{equation}
a=a^{\rm ref},
\qquad
b=b^{\rm ref},
\qquad
E_{\gamma}=E_{\Gamma}=E.
\end{equation}
The first two conditions in Eq. (43) seem quite natural,
whereas the last equality imposes rather strong restrictions
on the degradation process.
In what follows, it will be shown that the latter condition is
fulfilled for some polymers and is violated for others.

\subsection{Fitting of observations on poly($\alpha$--methyl\-styrene)}

We begin with the analysis of observations for
thermo-oxidative degradation on PAMS with the
initial mass-average molecular weight
$M_{\rm w}(0)=9.0$ kg/mol and the polydispersity index
$D=2$ reported by Sterling et al. \cite{SKM01}.
These data demonstrate that the time--temperature superposition
principle is satisfied for the degradation process
with a high level of accuracy.

In each test, a specimen was dissolved in 1,2,4-trichloro\-benzene,
heated to a fixed reaction temperature
$T$ in the range from 110 to 150~$^{\circ}$C,
and di-tert-butyl peroxide was added to a reaction vessel
with a constant flow rate under stirring.
Specimens for GPC (gel permeation chromatography) analysis
were taken every 45 min.
For a detailed description of the experimental procedure,
see \cite{SKM01}.

Changes in the ratio of number-average molecular weights
$d_{\rm n}$ with time $t$ are depicted in Figure 9.
The experimental data at $T^{\rm ref}=110$~$^{\circ}$C are
presented without changes.
Observations at the other temperatures
($T=120$, 130, 140 and 150~$^{\circ}$C)
are shifted along the time-axis by appropriate amounts
$A$ that are determined from the condition that the experimental
data produce a smooth master-curve.

The parameter $A$ is plotted in Figure 10 versus temperature
$T$.
The experimental data are approximated by the dependence
[that follows from Eq. (42)]
\begin{equation}
\ln A=A_{0}-\frac{A_{1}}{T}
\end{equation}
with
\begin{equation}
A=\frac{\gamma_{0}}{\gamma_{0}^{\rm ref}},
\qquad
A_{0}=\frac{E}{R T^{\rm ref}},
\qquad
A_{1}=\frac{E}{R}.
\end{equation}
The coefficients $A_{k}$ ($k=0,1$) in Eq. (44) are determined
by the least-squares method.
Figure 10 demonstrates that Eq. (44) ensures quite acceptable
fit of the observations.
According to Eq. (45) and Figure 10, the activation energy reads
$E=86$ kJ/mol.
This value is in reasonable agreement with the activation
energy $E=112$ kJ/mol found by Brown and Wall \cite{BW58}
by applying a graphic method (whose accuracy is rather low due
to the numerical estimation of derivatives)
and in excellent accord with the activation energies
recently determined for other polymers by using
more sophisticated techniques
($E=88$ kJ/mol for low-density polyethylene
\cite{KKM02} and $E=98$ kJ/mol for polypropylene
\cite{CB97}).

To reduce the number of adjustable parameters in Eq. (35),
we set $a=0$, which means that we presume the rate of
fragmentation to be independent of chains' length.
To find the quantities $\gamma_{0}$, $b$ and $\eta$,
we use an algorithm similar to that employed in Section 5.
We fix some intervals $[0,\gamma_{\max}]$,
$[0,b_{\max}]$ and $[0,\eta_{\max}]$,
where the ``best-fit" parameters $\gamma_{0}$, $b$
and $\eta$ are assumed to be located,
and divide these intervals into $J$ subintervals by
the points $\gamma^{(i)}=i \Delta \gamma$,
$b^{(j)}=j \Delta b$ and $\eta^{(k)}=k \Delta \eta$
($i,j,k=1,\ldots,J-1$)
with $\Delta \gamma=\gamma_{\max}/J$,
$\Delta b=b_{\max}/J$ and $\Delta \eta=\eta_{\max}/J$.
For any triple $\{ \gamma^{(i)}, b^{(j)}, \eta^{(k)} \}$,
Eq. (35) with initial condition (33) is integrated numerically
by the Runge--Kutta method with $K=100$ and
the time-step $\Delta t=0.1$.
The best-fit parameters $\gamma_{0}$, $b$ and $\eta$ are
determined from the condition of minimum of the function
\begin{equation}
R=\sum_{t_{m}} \Bigl [ d_{\rm n}^{\;\rm exp}(t_{m})
-d_{\rm n}^{\;\rm num}(t_{m}) \Bigr ]^{2},
\end{equation}
where the sum is calculated over all times $t_{m}$
at which observations are presented in Figure 9,
$d_{\rm n}^{\;\rm exp}$ is the ratio of number-average
molecular weights measured in the tests,
and $d_{\rm n}^{\;\rm num}$ is given by Eqs. (7), (8)
and (31).
Figure 9 demonstrates good agreement between the
observations at various temperatures $T$
and the results of numerical simulation with
$\gamma_{0}=9.0\cdot 10^{-7}$, $b=1.5$ and $\eta=50.0$.

As the value of $\eta$ is rather large compared to unity,
we draw a conclusion from Figures 1, 2 and 9 that the effect
of annihilation of end- and side-groups on the degradation
kinetics of PAMS is substantial.
The observation that the rate of annihilation for PAMS
($\eta=50$) exceeds that for PS ($\eta=1$) by about two
orders of magnitude appears to be natural,
because PAMS chains differ from those of PS by the presence
of methyl groups attached to each tertiary backbone carbon
\cite{SKM01}.

\subsection{Fitting of observations on poly($L$--lactide)}

We proceed with the analysis of experimental data on
thermal degradation of PLLA reported by Yu et al. \cite{YHW03}.
PLLA samples with the initial mass-average molecular weight
$M_{\rm w}(0)=108$ kg/mol and the polydispersity index $D=1.5$
were heated to a required temperature $T$
(in the range between 180 to 220~$^{\circ}$C),
preserved at this temperature in a reaction vessel under
nitrogen atmosphere for given amounts of time (from 0.5 to 6.0 h),
subsequently removed from the vessel,
and their molecular weights were measured by GPC.
The ratio of the number-average molecular weights $d_{\rm n}$
is plotted versus exposure time $t$ in Figure 11.
This figure shows that the time--temperature superposition
principle is not valid for the degradation process
(the experimental data at $T=200$ and $T=220$~$^{\circ}$C
re-plotted in the semi-logarithmic scale cannot be superposed
by shifts along the time-axis with an acceptable level of
accuracy).
This conclusion may be explained by the fact that the activation
energies $E_{\gamma}$ and $E_{\Gamma}$
noticeably differ from each other.

To validate this result and to assess the difference between
the activation energies for fragmentation and annihilation,
we approximate each curve depicted in Figure 11 separately.
To reduce the number of adjustable parameters, we set
$a=0$ (which means that the fragmentation rate is assumed to
be independent of chains' length) and $b=2.5$
(this value is taken because the observations at $T=200$
and $T=220$~$^{\circ}$C depicted in Figure 11 resemble our
results of numerical simulation with $b=2.5$ reported in
Figure 5).
The fact that the rate of fragmentation in PLLA
specimens with relatively small molecular weights
(less than 300 kg/mol) is independent of chains' length
is confirmed by the experimental data reported by
Bywater and Black (see Figure 3 in \cite{BB65}).

The rates $\gamma_{0}$ and $\Gamma_{0}$ are determined
by using an algorithm similar to that described in Section 5.
We fix some intervals $[0,\gamma_{\max}]$ and $[0,\Gamma_{\max}]$,
where the ``best-fit" parameters $\gamma_{0}$ and $\Gamma_{0}$
are assumed to be located,
and divide these intervals into $J$ subintervals by
the points $\gamma^{(i)}=i \Delta \gamma$ and
$\Gamma^{(j)}=j \Delta \Gamma$ ($i,j=1,\ldots,J-1$)
with $\Delta \gamma=\gamma_{\max}/J$ and
$\Delta \Gamma=\Gamma_{\max}/J$.
For any pair $\{ \gamma^{(i)}, \Gamma^{(j)} \}$,
Eq. (35) with initial condition (33) is integrated numerically
by the Runge--Kutta method with $K=100$ and
the time-step $\Delta t=0.1$.
The best-fit parameters $\gamma_{0}$ and $\Gamma_{0}$ are
determined from the condition of minimum of the cost function
(46).
Figure 11 reveals good agreement between the observations
at all three temperatures $T$ under consideration
and the results of numerical simulation.
The ratio $\eta$ determined by Eq. (34) ranges from
0.9 at $T=180$ to 11.0 at $T=200$ and to 15.0 at
$T=220$~$^{\circ}$C.
These values together with the results depicted in Figures
1 and 2 show that the effect of annihilation of end- and
side-groups on the degradation process is quite substantial,
and the influence of this process increases with temperature.

The adjustable parameters $\gamma_{0}$ and $\Gamma_{0}$
are plotted versus temperature $T$ in Figure 12.
The experimental data are approximated by the equations
similar to Eq. (44),
\begin{equation}
\ln A=A_{0}-\frac{A_{1}}{T},
\qquad
\ln B=B_{0}-\frac{B_{1}}{T}
\end{equation}
with $T^{\rm ref}=453$~K (180~$^{\circ}$C) and
\begin{eqnarray}
A=\frac{\gamma_{0}}{\gamma_{0}^{\rm ref}},
\qquad
A_{0}=\frac{E_{\gamma}}{R T^{\rm ref}},
\qquad
A_{1}=\frac{E_{\gamma}}{R},
\nonumber\\
B=\frac{\Gamma_{0}}{\Gamma_{0}^{\rm ref}},
\qquad
B_{0}=\frac{E_{\Gamma}}{R T^{\rm ref}},
\qquad
B_{1}=\frac{E_{\Gamma}}{R} .
\end{eqnarray}
Equations (47) follow from Eq. (42).
Figure 12 shows that Eq. (47) provides an acceptable
fit of the observations
(it should be noted some scatter of the experimental data).
However, the deviations of the data from Eq. (42)
are noticeably less pronounced that the difference between
curves 1 and 2 in Figure 12, which confirms our hypothesis
that fragmentation and annihilation of chains may be governed
by different processes at the micro-level.

The activation energy found for the fragmentation process,
$E_{\gamma}=91.9$ kJ/mol, is in good agreement
with the value of $E_{\gamma}$ determined for PAMS.
The fact that the activation energy for annihilation of
end- and side-groups ($E_{\Gamma}=223.9$ kJ/mol)
exceeds by about twice that for fragmentation of chains
implies that the annihilation process strongly affects
polymer degradation at elevated temperatures, whereas
its influence is of secondary impotence at relatively low
temperatures.
This conclusion provides a clue to explain our results
of numerical simulation on PS specimens: the temperature
$T=275$~$^{\circ}$C is relatively low for intensive breakage
of end- and side-groups in polystyrene, which implies that
the value $\eta=1.0$ found by matching the experimental
data is rather small.

\section{Thermo-gravimetric analysis}

It was revealed in the previous sections that
the model can correctly describe experimental data
on the evolution of molecular weights
with exposure time in conventional depolymerization tests.
It should be noted, however, that other kinetic models grounded on
either the fragmentation concept or the aggregation--fragmentation
theory can reproduce at least some of these observations
with a reasonable level of accuracy.
What cannot be described by the conventional models
(all of which imply the conservation law for the number of
segments in a network) is observations in thermo-gravimetric
experiments, where a decrease in a sample's mass is measured
as a function of exposure time at elevated temperatures.

Our aim now is to demonstrate that (i) the model grounded
on the fragmentation--annihilat\-ion concept can qualitatively
describe experimental data in thermo-gravimetric tests,
and (ii) the kinetic equations with adjustable parameters
determined by matching observations in GPC tests
can quantitatively predict results of thermo-gravimetric
analysis (TGA).

Two kinds of TGA tests are conventionally performed.
In experiments with a constant rate of heating $q=dT/dt$,
mass $\bar{m}$ of a specimen is measured as a function of
exposure time $t$ and the ratio
\[
d(t)=\frac{\bar{m}(t)}{\bar{m}(0)}
\]
of the current mass $\bar{m}(t)$ to the initial mass $\bar{m}(0)$
is plotted as a function of current temperature $T(t)$.
The graph $d(T)$ demonstrates the following features:
(i) $d$ remains practically constant and equal unity up to
some temperature $T_{\rm onset}$, at which the
degradation process starts,
(ii) $d$ sharply decreases with temperature (within
an interval of temperatures smaller than 100 K),
and (iii) $d$ practically vanishes at higher temperatures.
The interval of pronounced changes in $d$ is characterized
by the depolymerization temperature $T_{\rm depol}$ that
describes the life-time of a polymer and equals the temperature
at which the mass loss reaches 5 \%.

In GPA tests with a constant temperature $T$,
a specimen is rapidly heated to the required temperature,
which remains constant during the experimental procedure,
and the sample's mass is measured as a function of exposure
time $t$.
The function $d(t)$ monotonically decreases from $d(0)=1$
and tends to zero with the growth of exposure time $t$.
The decrease in mass at a given temperature $T$ is
characterized by the half-life time $t_{\rm hl}$ that
is determined as the exposure time at which the sample
loses half of its initial mass.

To compare predictions of the model with the experimental
data available in the literature, we perform numerical
integration of Eqs. (33) and (35) with $K=100$ and
$\Delta t=1.0\cdot 10^{-3}$ [these values guarantee that
the evolution of the first moments $M_{m}(t)$ ($m=0,1,2$)
is independent of the integration algorithm]
for poly($\alpha$-methylstyrene) with $a=0.0$, $b=1.5$
and $\eta=50.0$ (these values are found by matching
the experimental data presented in Figure 9).
The effect of temperature $T$ on the rate of fragmentation
$\gamma_{0}$ is described by Eqs. (44) and (45) with
$\gamma_{0}^{\rm ref}=9.0\cdot 10^{-7}$,
$A_{0}=26.86$ and $A_{1}=10344.0$
(these values are determined by matching the observations
depicted in Figure 10).
To take into account the influence of the initial
molecular weight $M_{\rm w}(0)$, we use Eq. (41) with
$M_{\rm w}^{\rm ref}(0)=9.0$ kg/mol.
The ratio of the current mass $\bar{m}(t)$ to the initial mass
$\bar{m}(0)$ is determined by Eq. (31), where the moment
$M_{1}(t)$ is given by Eq. (12).

Integration of Eq. (35) is carried out for
$M_{\rm w}(0)=4.0$ kg/mol and $M_{\rm w}=680.0$ kg/mol
(in the latter case, a correction factor of 1.5
is introduced in Eq. (41) in accord with the experimental data
reported in Figure 3 of \cite{BB65})
for the first kind of thermo-gravimetric tests with
$q=10.0$ K/min and for $M_{\rm w}=680.0$ kg/mol for the
other kind of experiments at the temperature $T=560$~K.
The results of numerical analysis are presented in Figures
13 and 14.
These figures demonstrate that the shapes of the curves
$d(T)$ and $d(t)$ are similar to those reported in numerous
experimental studies.

To show that our results of numerical simulation provide
quantitative coincidence with observations, we depict
(in Figure 13) the onset temperature $T_{\rm onset}$ for
degradation of PAMS with the low molecular weight
$M_{\rm w}(0)=4.0$ kg/mol \cite{BG03},
and the depolymerization temperature $T_{\rm depol}$
for PAMS with the high molecular weight $M_{\rm w}=680.0$
kg/mol \cite{KS99}.
For the same purpose, we plot the boundaries of the interval
for the half-life time $t_{\rm hl}$ of PAMS at $T=560$~K
reported in \cite{KS99}.
Striking similarity is observed between the experimental data
and the model predictions.

\section{Conclusions}

A kinetic model is analyzed for thermal degradation of polymers.
The governing relations are based on the
fragmentation--annihilation concept.
According to this approach, thermal degradation
reflects two processes at the micro-level:
(i) binary fragmentation of macromolecules
and (ii) scission of end- and side-groups and their
annihilation (diffusion and evaporation) from the network
of chains.
The evolution equations involve four material constants
that are determined by fitting experimental data for
polystyrene, poly($\alpha$--methylstyrene) and poly(L-lactide).

Two explicit solutions of the integro-differential equations
are found by using the method of generating function and the
Charlesby method.
These solutions (corresponding to special values of
adjustable parameters) serve as a basis for numerical analysis.
Numerical simulation of the kinetic equations is performed
to study the effect of material constants on the evolution
of mass-average and number-average molecular weights of
polymers with exposure time.

With reference to available experimental data, two superposition
principles (time--temperat\-ure and time--molecular weight) are
formulated, and restrictions on material constants are found
that guarantee their validity.
These limitations allow the number of adjustable parameters
to be reduced in order to make the approximation of observations
more reliable.

It is revealed that the fragmentation--annihilation concept
not only can be applied to describe experimental data
on changes in the molecular weight, but can also be used to
predict (qualitatively, and in some cases quantitatively)
the mass decrease observed in TGA tests.

The following conclusions may be drawn from our analysis
of experimental data on PS, PAMS and PLLA:
\begin{itemize}
\item
The conventional assumption about the effect of chains'
length ($a>0$) on the fragmentation process is excessive.
It results in a noticeable complication of the theoretical
analysis, but does not improve the quality of fitting.

\item
The account of the influence of chains' length on the rate
of annihilation is essential.
The parameter $b$ weakly affects changes in the molecular
weight with time at $b\in (0,1)$, but its effect becomes
important at higher values of the exponent $b$.

\item
The kinetics of polymer degradation is noticeably affected
by the annihilation process when two conditions are fulfilled:
(i) the number of end- and side-groups in polymer chains
is rather large,
and (ii) the rate of annihilation exceeds the rate of
fragmentation of chains by (at least) an order of magnitude.
Our results of numerical simulation demonstrate that the latter
condition is satisfied at relatively high temperatures.
\end{itemize}

\renewcommand{\theequation}{A-\arabic{equation}}
\setcounter{equation}{0}
\section*{Appendix}

The aim of this section is to derive an explicit formula
for the evolution of the second moment $M_{2}$
with time $t$, when the fragmentation process is described
by the conventional equation
\begin{equation}
\frac{\partial n}{\partial t}(t,x) = -\gamma_{0} n(t,x)
+2\gamma_{0} \int_{x}^{\infty} n(t,y) dy
\end{equation}
with an arbitrary initial condition (11).
Equation (A-1) coincides with Eq. (11) where the annihilation
process is disregarded ($\Gamma_{0}=0$).
The derivation is presented as an appendix, because we use
a conventional approach (the method of generating functions)
for the study of Eq. (A-1).
To the best of our knowledge, the final result is, however, novel.

We begin with the analysis of the moments $M_{0}(t)$ and
$M_{1}(t)$.
It follows from Eq. (15) that these functions are governed
by the differential equation
\begin{equation}
\frac{dM_{0}}{dt}(t)=\gamma_{0} M_{1}(t),
\qquad
\frac{dM_{1}}{dt}(t)=0.
\end{equation}
Integrating Eq. (A-2) and bearing in mind that
\[
M_{0}(0)=1
\]
[this equality follows from Eqs. (1), (8) and (12)],
we obtain
\begin{equation}
M_{0}(t)=1+\gamma_{0} M_{1}(0)t,
\qquad
M_{1}(t)=M_{1}(0).
\end{equation}
We now multiply Eq. (A-1) by $\exp(-zx)$, where $z$ is a new
variable, and integrate the result over $x$ from zero to
infinity.
Introducing the notation [the Laplace transform of
the function $n(t,x)$]
\begin{equation}
\hat{n}(t,z)=\int_{0}^{\infty} n(t,x)\exp(-zx)dx,
\end{equation}
we find that
\begin{equation}
\frac{\partial \hat{n}}{\partial t}(t,z) = -\gamma_{0}
\int_{0}^{\infty} x n(t,x) \exp(-z x) dx
+2\gamma_{0} \int_{0}^{\infty} \exp(-z x) dx
\int_{x}^{\infty} n(t,y) dy .
\end{equation}
According to Eq. (A-4), the first integral on the right-hand side
of Eq. (A-5) reads
\[
\int_{0}^{\infty} x n(t,x) \exp(-z x) dx
=-\frac{\partial \hat{n}}{\partial z}(t,z).
\]
To transform the other integral, we change the order of
integration and use Eqs. (12) and (A-4),
\begin{eqnarray*}
&& \int_{0}^{\infty} \exp(-z x) dx \int_{x}^{\infty} n(t,y) dy
=\int_{0}^{\infty} n(t,y) dy\int_{0}^{y} \exp(-z x) dx
\nonumber\\
&&=\frac{1}{z} \int_{0}^{\infty} \Bigl [1-\exp(-z y)\Bigr ]n(t,y)
dy =\frac{1}{z} \Bigl [ M_{0}(t)-\hat{n}(t,z) \Bigr ].
\end{eqnarray*}
Substitution of these expressions into Eq. (A-5) results
in the partial differential equation
\begin{equation}
\frac{\partial \hat{n}}{\partial t}(t,z)
=\gamma_{0} \frac{\partial \hat{n}}{\partial z}(t,z)
+\frac{2\gamma_{0}}{z} \Bigl [ M_{0}(t)-\hat{n}(t,z) \Bigr ] .
\end{equation}
Introducing the function
\begin{equation}
\Phi(t,z)=\hat{n}(t,z)-M_{0}(t),
\end{equation}
we present Eq. (A-6) in the form
\[
\frac{\partial \Phi}{\partial t}(t,z)+\frac{dM_{0}}{dt}(t)
=\gamma_{0} \frac{\partial \Phi}{\partial z}(t,z)
-\frac{2\gamma}{z} \Phi(t,z).
\]
It follows from this equality and Eq. (A-2) that
\begin{equation}
\frac{\partial \Phi}{\partial t}(t,z)
-\gamma_{0} \frac{\partial \Phi}{\partial z}(t,z)
=-\frac{2\gamma_{0}}{z} \Phi(t,z) -\gamma_{0} M_{1}(0).
\end{equation}
We search a solution of Eq. (A-8) in the form
$\Phi=\Psi(x_{1},x_{2})$, where
\begin{equation}
x_{1}=t+\frac{z}{\gamma_{0}},
\qquad
x_{2}=z.
\end{equation}
In the new notation, Eq. (A-8) reads
\begin{equation}
\frac{\partial \Psi}{\partial x_{2}}(x_{1},x_{2})
=M_{1}(0) +\frac{2}{x_{2}} \Psi(x_{1},x_{2}).
\end{equation}
An advantage of Eq. (A-10) is that it includes $x_{1}$ as a
parameter.
We now introduce the function $\Psi_{1}(x_{1},x_{2})$ by the formula
\begin{equation}
\Psi(x_{1},x_{2})=x_{2}^{2} \Psi_{1}(x_{1},x_{2}).
\end{equation}
Equations (A-10) and (A-11) imply that this function satisfies
the equation
\begin{equation}
\frac{\partial \Psi_{1}}{\partial x_{2}}(x_{1},x_{2})
=\frac{M_{1}(0)}{x_{2}^{2}}.
\end{equation}
The general solution of Eq. (A-12) reads
\[
\Psi_{1}(x_{1},x_{2})=-\frac{M_{1}(0)}{x_{2}}+F(x_{1}),
\]
where $F$ is an arbitrary smooth function of $x_{1}$.
Returning to the initial notation with the help of Eqs. (A-7),
(A-9) and (A-11), we obtain
\begin{equation}
\hat{n}(t,z)=M_{0}(t)-M_{1}(0)z+F\Bigl (t+\frac{z}{\gamma_{0}}\Bigr )z^{2}.
\end{equation}
It follows from Eqs. (11) and (A-4) that
\begin{equation}
\hat{n}(0,z)=\hat{n}_{0}(z),
\qquad
\hat{n}_{0}(z)=\int_{0}^{\infty} \exp(-zx) n_{0}(x) dx.
\end{equation}
Setting $t=0$ in Eq. (A-13) and using Eq. (A-14), we find that
\begin{equation}
F(z)=\frac{1}{(\gamma_{0} z)^{2}}\Bigl [ \hat{n}_{0}(\gamma_{0} z)
-M_{0}(0)+M_{1}(0) \gamma_{0} z\Bigr ].
\end{equation}
Equations (12) and (A-4) imply that
\begin{equation}
M_{2}(t)=\frac{\partial^{2} \hat{n}}{\partial z^{2}}(t,0).
\end{equation}
Differentiation of Eq. (A-13) with respect to $z$ results in
\[
\frac{\partial^{2} \hat{n}}{\partial z^{2}}(t,0)=2F(t).
\]
Combining this equality with Eqs. (A-15) and (A-16),
we arrive at the formula
\begin{equation}
M_{2}(t)=\frac{2}{(\gamma_{0} t)^{2}}\Bigl [
\hat{n}_{0}(\gamma_{0} t) -M_{0}(0)+M_{1}(0) \gamma_{0} t\Bigr ].
\end{equation}
Equations (A-3) and (A-17) provide explicit formulas for the
evolution of three first moments $M_{m}$ driven by thermal
degradation of a polymer with an arbitrary initial distribution
of chain lengths.

\newpage

\subsection*{List of figures}

\begin{description}

\item{
{\bf Figure 1:}
The ratio of number-average molecular weights
$d_{\rm n}$ versus dimensionless time $t$.
Curves: results of numerical simulation for
$a=0.0$, $b=1.0$ and $\eta=0.0$, 1.0, 10.0,
100.0 and 200.0, from top to bottom, respectively}
\vspace*{1 mm}

\item{
{\bf Figure 2:}
The ratio of mass-average molecular weights
$d_{\rm w}$ versus dimensionless time $t$.
Curves: results of numerical simulation for
$a=0.0$, $b=1.0$ and $\eta=0.0$, 1.0, 10.0,
100.0 and 200.0, from top to bottom, respectively}
\vspace*{1 mm}

\item{
{\bf Figure 3:}
The ratio of number-average molecular weights
$d_{\rm n}$ versus dimensionless time $t$.
Curves: results of numerical simulation for
$b=1.0$, $\eta=10.0$ and $a=0.3$, 0.6, 0.9
and 1.2, from top to bottom, respectively}
\vspace*{1 mm}

\item{
{\bf Figure 4:}
The ratio of mass-average molecular weights
$d_{\rm w}$ versus dimensionless time $t$.
Curves: results of numerical simulation for
$b=1.0$, $\eta=10.0$ and $a=0.3$, 0.6, 0.9
and 1.2, from top to bottom, respectively}
\vspace*{1 mm}

\item{
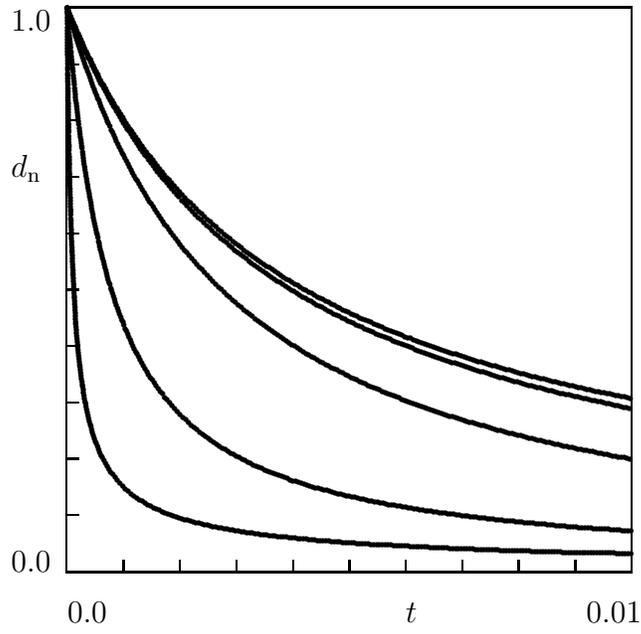
{\bf Figure 5:}
The ratio of number-average molecular weights
$d_{\rm n}$ versus dimensionless time $t$.
Curves: results of numerical simulation for
$a=0.2$, $\eta=10.0$ and $b=0.0$, 1.0, 1.5, 2.0
and 2.5, from top to bottom, respectively}
\vspace*{1 mm}

\item{
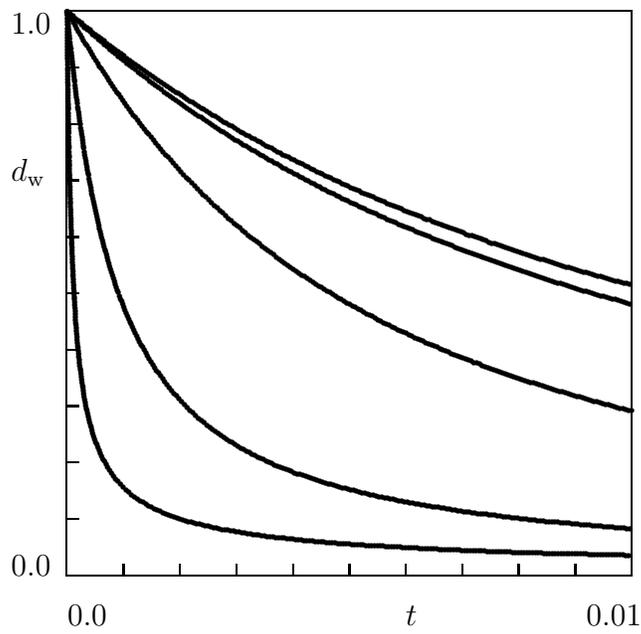
{\bf Figure 6:}
The ratio of mass-average molecular weights
$d_{\rm w}$ versus dimensionless time $t$.
Curves: results of numerical simulation for
$a=0.2$, $\eta=10.0$ and $b=0.0$, 1.0, 1.5, 2.0
and 2.5, from top to bottom, respectively}
\vspace*{1 mm}

\item{
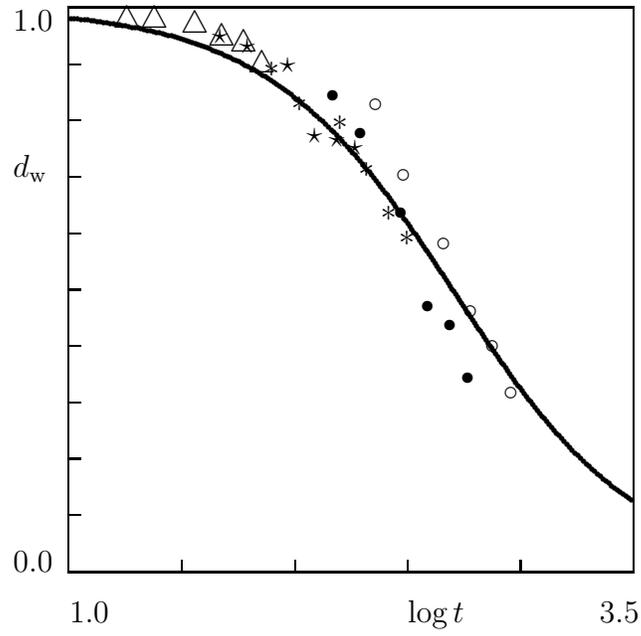
{\bf Figure 7:}
The ratio of mass-average molecular weights
$d_{\rm w}$ versus time $t$ min.
Symbols: treatment of observations on PS
with the initial molecular weights $M_{\rm w}(0)=26$
(triangles), 110 (stars), 210 (asterisks),
330 (filled circles) and 930  kg/mol (unfilled circles).
Solid line: results of numerical simulation}
\vspace*{1 mm}

\item{
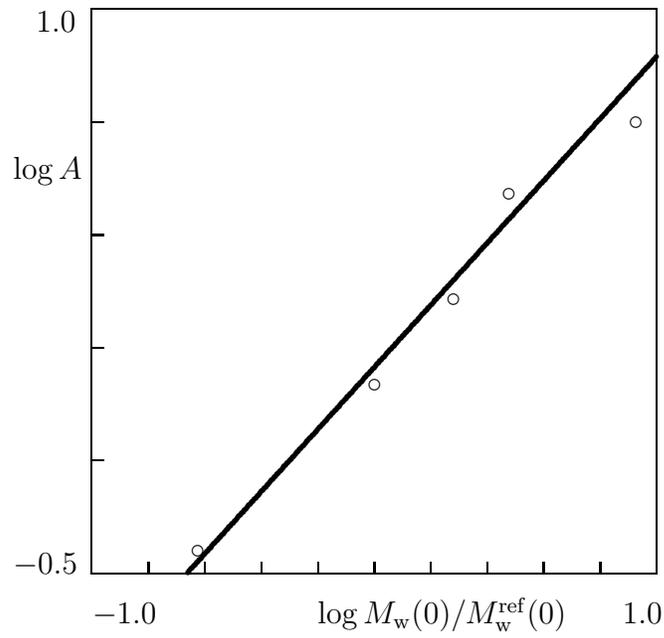
{\bf Figure 8:}
The shift factor $A$ versus the ratio of
mass-average molecular weights.
Circles: treatment of observations on PS.
Solid line: approximation of the experimental data
by Eq. (40) with $A_{0}=0.05$ and $A_{1}=0.83$}
\vspace*{1 mm}

\item{
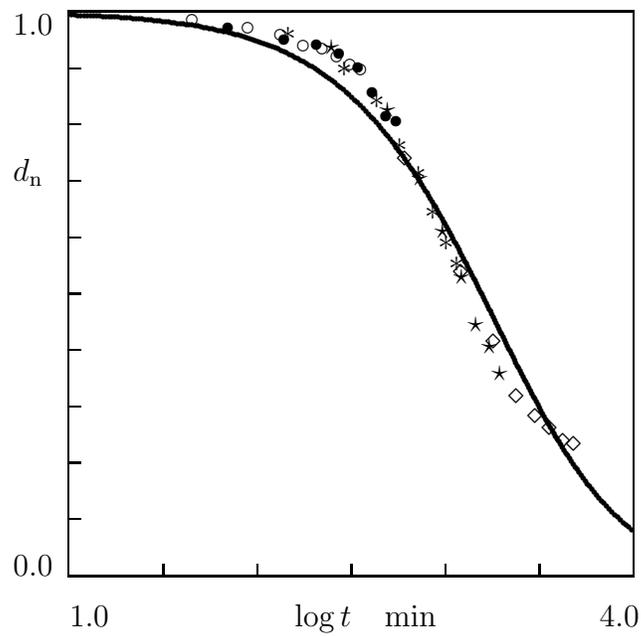
{\bf Figure 9:}
The ratio of number-average molecular weights
$d_{\rm n}$ versus time $t$.
Symbols: treatment of observations on PAMS
at the temperatures $T=110$ (unfilled circles),
$T=120$ (filled circles),
$T=130$ (asterisks),
$T=140$ (stars) and $T=150$~$^{\circ}$C (diamonds).
Solid line: results of numerical simulation}
\vspace*{1 mm}

\item{
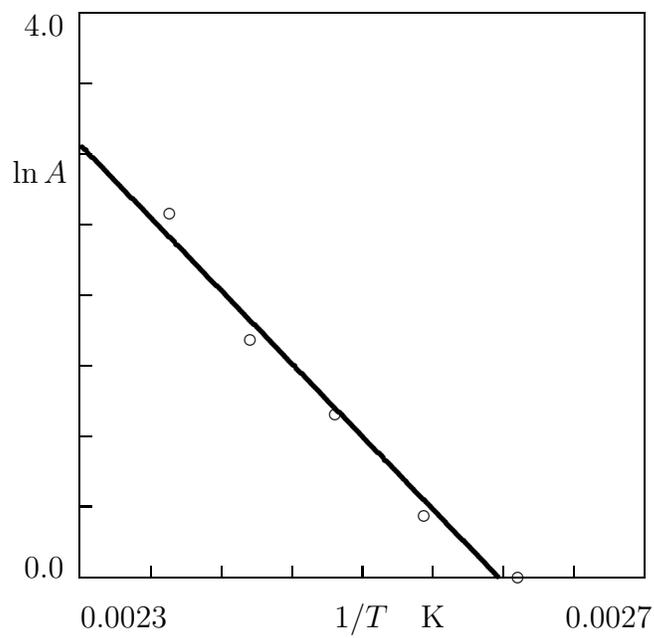
{\bf Figure 10:}
The shift factor $A$ versus temperature $T$.
Circles: treatment of observations on PAMS.
Solid line: approximation of the experimental data
by Eq. (44) with $A_{0}=26.86$ and $A_{1}=1.03\cdot 10^{4}$}
\vspace*{1 mm}

\item{
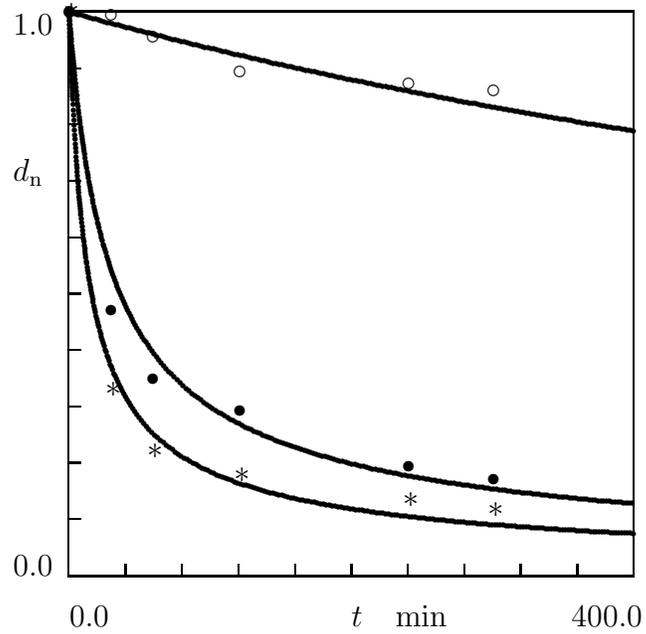
{\bf Figure 11:}
The ratio of number-average molecular weights
$d_{\rm n}$ versus time $t$.
Symbols: experimental data on PLLA.
Solid lines: results of numerical simulation}
\vspace*{1 mm}

\item{
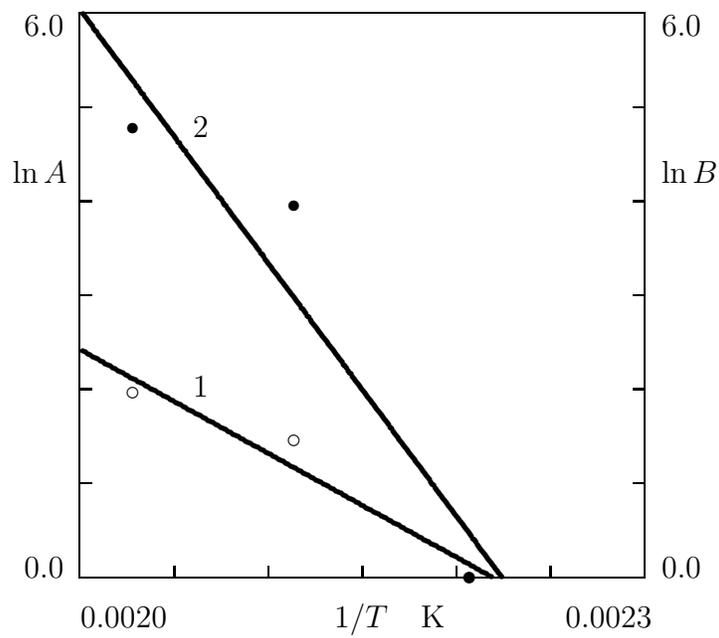
{\bf Figure 12:}
The shift factors $A$ (unfilled circles)
and $B$ (filled circles) versus temperature $T$.
Circles: treatment of observations on PLLA.
Solid lines: approximation of the experimental data
by Eq. (47).
Curve 1: $A_{0}=24.53$, $A_{1}=1.11\cdot 10^{4}$.
Curve 2: $B_{0}=59.90$, $B_{1}=2.69\cdot 10^{4}$}
\vspace*{1 mm}

\item{
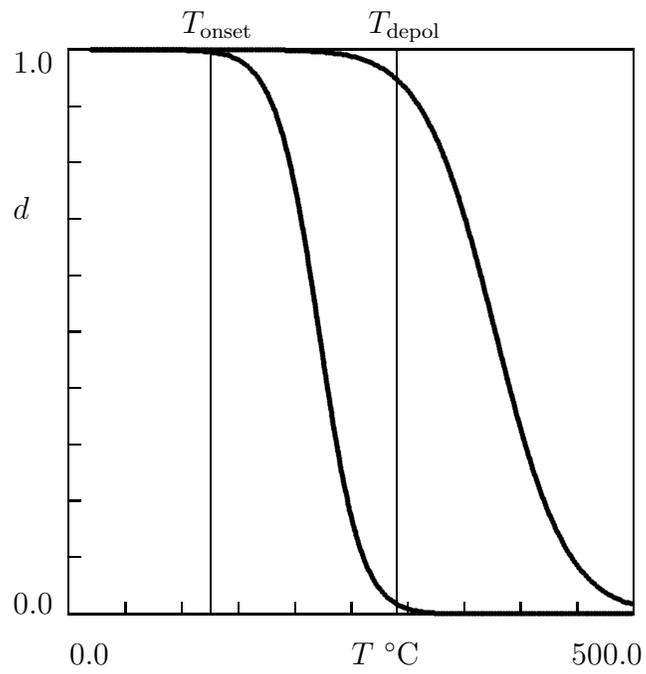
{\bf Figure 13:}
The ratio $d$ of a sample's mass to its initial
mass versus temperature $T$.
Solid lines: results of numerical simulation for PAMS
under heating with the rate $q=10.0$ K/min.
Curve 1: $M_{\rm w}(0)=4.0$ kg/mol.
Curve 2: $M_{\rm w}(0)=680.0$ kg/mol.
Vertical lines: the onset temperature $T_{\rm onset}$
for thermal degradation of the low-molecular weight PAMS
and the depolymerization temperature $T_{\rm depol}$
for the high-molecular weight PAMS}
\vspace*{1 mm}

\item{
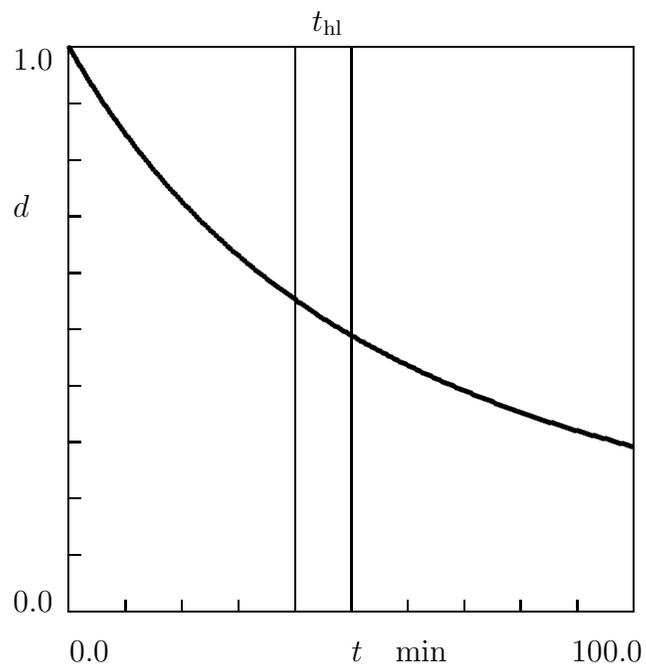
{\bf Figure 14:}
The ratio $d$ of a sample's mass to its initial mass
versus exposure time $t$ at the temperature $T=560$~K.
Solid line: results of numerical simulation for PAMS.
Vertical lines denote the interval of time where a specimen
loses a half of its initial mass}
\end{description}
\vspace*{100 mm}

\setlength{\unitlength}{0.75 mm}
\begin{figure}[tbh]
\begin{center}

\end{center}
\vspace*{10 mm}

\caption{}
\end{figure}

\begin{figure}[tbh]
\begin{center}
%
\end{center}
\vspace*{10 mm}

\caption{}
\end{figure}

\begin{figure}[tbh]
\begin{center}
%
\end{center}
\vspace*{10 mm}

\caption{}
\end{figure}

\begin{figure}[tbh]
\begin{center}
%
\end{center}
\vspace*{10 mm}

\caption{}
\end{figure}

\end{document}